\documentclass[aps,rpl,preprint,groupedaddress]{revtex4}
\begin{document}

\title{Three-qubit Thermal Entanglement via Entanglement Swapping on Two-qubit Heisenberg $XY$ chains}

\author{Zi Chong Kao}
\affiliation{Hwa Chong Institution, 673 Bukit Timah Road, Singapore 269735, Singapore}

\author{Jezreel Ng}
\affiliation{Hwa Chong Institution, 673 Bukit Timah Road, Singapore 269735, Singapore}

\author{Ye Yeo}
\affiliation{Department of Physics, National University of Singapore, 10 Kent Ridge Crescent, Singapore 119260, Singapore}

\begin{abstract}
In this paper we consider the generation of a three-qubit GHZ-like thermal state by applying the entanglement swapping scheme of Zukowski {\it et al.} [Ann. N. Y. Acad. Sci. {\bf 755}, 91 (1995)] to three pairs of two-qubit Heisenberg $XY$ chains.  The quality of the resulting three-qubit entanglement is studied by analyzing the teleportation fidelity, when it is used as a resource in the teleportation protocol of Karlsson {\it et al.}[Phys. Rev. A {\bf 58}, 4394 (1998)].  We show that even though thermal noise in the original two-qubit states is amplified by the entanglement swapping process, we are still able to achieve nonclassical fidelities for the anisotropic Heisenberg $XY$ chains at finitely higher and higher temperatures by adjusting the strengths of an external magnetic field.  This has a positive implication on the solid-state realization of a quantum computer.
\end{abstract}

\maketitle

\section{Introduction}
Algorithms such as quantum factoring \cite{Shor} and quantum search \cite{Grover} illustrate the great theoretical promise of quantum computers \cite{Nielsen}.  Consequently, many designs for the practical implementation of such devices have been proposed.  Among these is one using quantum dots (localized electron spins) as quantum bits (qubits \cite{Schumacher}), which has its roots in solid state physics \cite{Loss, Imamoglu}.  The recent breakthroughs in the experimental physics of double quantum dot \cite{Chen} show that this proposal is especially promising.  Mathematically, a two-qubit Heisenberg $XY$ chain at finite temperatures can serve as an effective model of a double quantum dot.  The state of a two-qubit Heisenberg $XY$ chain at thermal equilibrium (temperature $T$) is described by $\chi = e^{-\beta H}/Z$, where $H$ is the Hamiltonian, $Z = {\rm tr}e^{-\beta H}$ is the partition function, and $\beta = 1/kT$, where $k$ is the Boltzmann's constant.  The entanglement associated with the thermal state $\chi$ is referred to as the thermal entanglement \cite{Arnesen}.  

In a very important 1999 paper, Ref.\cite{Gottesman}, Gottesman and Chuang showed that single qubit operations, Bell-basis measurements and Greenberger-Horne-Zeilinger (GHZ) states \cite{Greenberger} are sufficient to construct a universal quantum computer.  The generation of GHZ states therefore becomes an important problem.  Slightly earlier, Zukowski {\it et al.} \cite{Zukowski} and Bose {\it et al.} \cite{Bose} had shown how to obtain GHZ states from say ``simpler'' Bell pairs via entanglement swapping.  Motivated by these experimental and theoretical developments, in this work, we consider as a first step the generation of a three-qubit GHZ-like thermal state from three pairs of two-qubit Heisenberg $XY$ chains using the technique of entanglement swapping \cite{Comment}.  Since quantum teleportation \cite{Bennett} lies at the heart of Gottesman and Chuang's argument, we determine the quality of the resulting three-qubit entanglement by analyzing the teleportation fidelity, when it is used as a resource in the teleportation protocol of Karlsson {\it et al.} \cite{Karlsson}, a generalization of the standard teleportation protocol of Bennett {\it et al.} \cite{Bennett} to one using entangled three-qubit resource.

This paper is thus organized as follows.  In Section II, we give the Hamiltonian for a two-qubit Heisenberg $XY$ chain in the presence of an external magnetic field.  Properties such as concurrence \cite{Hill, Wootters} and fully-entangled fraction \cite{Popescu, Horodecki} of the associated thermal state are presented in Section III.  In Sections IV and V, we briefly describe the entanglement swapping scheme of Zukowski {\it et al.} and the teleportation scheme of Karlsson {\it et al.} respectively.  They are cast into forms suitable for our considerations here.  We give the resulting three-qubit thermal state in Section IV and the formula for calculating the correspoding teleportation fidelity in Section V.  These set the stage necessary for the presentation of our results in Section VI.  First, using the isotropic two-qubit Heisenberg $XX$ chains, we establish that the entanglement swapping procedure amlpifies noise originally present in the two-qubit thermal states.   Next, we show that despite this, one is still able to achieve nonclassical fidelities for the anisotropic Heisenberg $XY$ chains at finitely higher and higher temperatures by adjusting the strengths of an external magnetic field.  In the concluding Section VII, we summarize and discuss two possible research problems related to our results here.


\section{Two qubit Heisenberg $XY$ chain}
The Hamiltonian $H$ for a two-qubit Heisenberg $XY$ chain in an external magnetic field $B_m \equiv \eta J$ ($\eta$ is a real number) along the $z$ axis is \cite{Kamta}
\begin{equation}
H = \frac{1}{2}(1 + \gamma)J\sigma^1_A \otimes \sigma^1_B + \frac{1}{2}(1 - \gamma)J\sigma^2_A \otimes \sigma^2_B + \frac{1}{2}B_m(\sigma^3_A \otimes \sigma^0_B + \sigma^0_A \otimes \sigma^3_B),
\end{equation}
where $\sigma^0_{\alpha}$ is the identity matrix and $\sigma^i_{\alpha}$ $(i = 1, 2, 3)$ are the Pauli matrices at site $\alpha = A, B$.  The parameter $-1 \leq \gamma \leq 1$ measures the anisotropy of the system and equals $0$ for the isotropic $XX$ model \cite{Wang1} and $\pm 1$ for the Ising model \cite{Gunlycke}.  $(1 + \gamma)J$ and $(1 - \gamma)J$ are real coupling constants for the spin interaction.  The chain is said to be antiferromagnetic for $J > 0$ and ferromagnetic for $J < 0$.

The eigenvalues and eigenvectors of $H$ are given by
\begin{eqnarray}
H|\Phi^0\rangle_{AB} & = & {\cal B} |\Phi^0\rangle_{AB}, \nonumber \\
H|\Phi^1\rangle_{AB} & = & J        |\Phi^1\rangle_{AB},\ \nonumber \\
H|\Phi^2\rangle_{AB} & = & -J       |\Phi^2\rangle_{AB},\ \nonumber \\
H|\Phi^3\rangle_{AB} & = & -{\cal B}|\Phi^3\rangle_{AB},
\end{eqnarray}
where ${\cal B} \equiv \sqrt{B^2_m + \gamma^2J^2} = \sqrt{\eta^2 + \gamma^2}J$,
\begin{eqnarray}
|\Phi^0\rangle_{AB} & = & \frac{1}{\sqrt{({\cal B} + B_m)^2 + \gamma^2J^2}}
[({\cal B} + B_m)|00\rangle_{AB} + \gamma J|11\rangle_{AB}], \\
|\Phi^1\rangle_{AB} & = & \frac{1}{\sqrt{2}}[|01\rangle_{AB} + |10\rangle_{AB}], \\
|\Phi^2\rangle_{AB} & = & \frac{1}{\sqrt{2}}[|01\rangle_{AB} - |10\rangle_{AB}], \\
|\Phi^3\rangle_{AB} & = & \frac{1}{\sqrt{({\cal B} - B_m)^2 + \gamma^2J^2}}
[({\cal B} - B_m)|00\rangle_{AB} - \gamma J|11\rangle_{AB}].
\end{eqnarray}
While 
$|\Phi^1\rangle_{AB} = \frac{1}{\sqrt{2}}[|01\rangle_{AB} + |10\rangle_{AB}] \equiv |\Psi^1_{Bell}\rangle_{AB}$ 
and 
$|\Phi^2\rangle_{AB} = \frac{1}{\sqrt{2}}[|01\rangle_{AB} - |10\rangle_{AB}] \equiv |\Psi^2_{Bell}\rangle_{AB}$ 
are maximally entangled Bell states, $|\Phi^0\rangle_{AB}$ and $|\Phi^3\rangle_{AB}$ are, so long as $\gamma \not= 0$, nonmaximally entangled states (see Eq.(18)).  In fact, Eqs.(3) and (6) reduce to the maximally entangled Bell states
$\frac{1}{\sqrt{2}}[|00\rangle_{AB} + |11\rangle_{AB}] \equiv |\Psi^0_{Bell}\rangle_{AB}$ 
and 
$\frac{1}{\sqrt{2}}[|00\rangle_{AB} - |11\rangle_{AB}] \equiv |\Psi^3_{Bell}\rangle_{AB}$ 
respectively, only when $B_m = 0$ (or $\eta = 0$).  For $\gamma = 0$, we have $|\Phi^0\rangle_{AB} = |00\rangle_{AB}$ and $|\Phi^3\rangle_{AB} = |11\rangle_{AB}$, product eigenstates with eigenvalues $B_m$ and $-B_m$ respectively \cite{Wang1}.

\section{Thermal state, concurrence and fully entangled fraction}
For the above system in thermal equilibrium at temperature $T$, its state is described by the density operator
\begin{equation}
\chi_{AB} = \frac{1}{Z}[e^{-\beta{\cal B}}|\Phi^0\rangle_{AB}\langle\Phi^0| + e^{-\beta J}|\Phi^1\rangle_{AB}\langle\Phi^1| + e^{\beta J}|\Phi^2\rangle_{AB}\langle\Phi^2| + e^{\beta{\cal B}}|\Phi^3\rangle_{AB}\langle\Phi^3|],
\end{equation}
where the partition function $Z = 2\cosh\beta{\cal B} + 2\cosh\beta J$, the Boltzmann's constant $k \equiv 1$ from hereon, and $\beta = 1/T$.  To quantify the amount of entanglement associated with $\chi_{AB}$, we consider the concurrence \cite{Hill, Wootters} 
${\cal C}[\chi_{AB}] \equiv \max\{\lambda_1 - \lambda_2 - \lambda_3 - \lambda_4,\ 0\}$, 
where $\lambda_k$ $(k = 1, 2, 3, 4)$ are the square roots of the eigenvalues in decreasing order of magnitude of the spin-flipped density-matrix operator 
$R_{AB} = \chi_{AB}(\sigma^2_A \otimes \sigma^2_B)\chi^*_{AB}(\sigma^2_A \otimes \sigma^2_B)$, 
the asterisk indicates complex conjugation.  After some straightforward algebra, we obtain
\begin{eqnarray}
\lambda_1 & = & \frac{1}{Z}e^{\beta J}, \\
\lambda_2 & = & \frac{1}{Z}e^{-\beta J}, \\
\lambda_3 & = & \frac{1}{Z}\sqrt{1 + \frac{2\gamma^2J^2}{{\cal B}^2}\sinh^2\beta{\cal B} + \frac{2\gamma J}{{\cal B}}\sqrt{1 + \frac{\gamma^2J^2}{{\cal B}^2}\sinh^2\beta{\cal B}}\sinh\beta{\cal B}}, \\
\lambda_4 & = & \frac{1}{Z}\sqrt{1 + \frac{2\gamma^2J^2}{{\cal B}^2}\sinh^2\beta{\cal B} - \frac{2\gamma J}{{\cal B}}\sqrt{1 + \frac{\gamma^2J^2}{{\cal B}^2}\sinh^2\beta{\cal B}}\sinh\beta{\cal B}}.
\end{eqnarray}
The concurrence derived from Eqs.(8) - (11) is invariant under the substitutions $\eta \longrightarrow -\eta$, $\gamma \longrightarrow -\gamma$, and $J \longrightarrow -J$.  The same applies to the fully entangled fraction ${\cal F}[\chi_{AB}]$. Therefore, we restrict our considerations to $\eta \geq 0$, $0 \leq \gamma \leq 1$, and $J > 0$ \cite{Yeo3}:
\begin{eqnarray}
{\cal F}[\chi_{AB}] & \equiv & \max_{i = 0, 1, 2, 3}\{{_{AB}}\langle\Psi^i_{Bell}|\chi_{AB}|\Psi^i_{Bell}\rangle_{AB}\} \nonumber \\
& = & \left\{\begin{array}{lll}
\frac{1}{Z}e^{\beta J}                                                                & {\rm for} & \sqrt{\eta^2 + \gamma^2} \leq 1, \\
\frac{1}{Z}\left(\cosh\beta{\cal B} + \frac{\gamma J}{\cal B}\sinh\beta{\cal B}\right)& {\rm for} & \sqrt{\eta^2 + \gamma^2} > 1.
\end{array}\right.
\end{eqnarray}
For a bipartite entangled state $\rho_{AB}$ to be useful for quantum teleportation we must have the fully entangled fraction ${\cal F}[\rho_{AB}] > \frac{1}{2}$.  $\rho_{AB}$ can then yield ``nonclassical'' teleportation fidelity $\Phi_{\max}$ ($> 2/3$) \cite{Popescu, Horodecki}.

Eq.(7) reduces to the following three possibilities in the zero-temperature limit, i.e., $\beta \longrightarrow \infty$, at which the system is in its ground state.
\begin{description}
\item{(a)} $\eta^2 + \gamma^2 < 1$:
\begin{eqnarray}
\chi_{AB} & = & 
\frac{1}{Z}[e^{\beta J}|\Phi^2\rangle_{AB}\langle\Phi^2| + e^{\beta{\cal B}}|\Phi^3\rangle_{AB}\langle\Phi^3|] \nonumber \\
& \longrightarrow & |\Phi^2\rangle_{AB}\langle\Phi^2|,
\end{eqnarray}
with $Z = e^{\beta J} + e^{\beta\cal B}$.  Eqs.(8) - (11) and (12) give ${\cal C}[\chi_{AB}] = 1$ and ${\cal F}[\chi_{AB}] = 1$ respectively, in agreement with the fact that $|\Phi^2\rangle_{AB}$ is a maximally entangled Bell state.
\item{(b)} $\eta^2 + \gamma^2 = 1$:
\begin{equation}
\chi_{AB} \longrightarrow \frac{1}{2}[|\Phi^2\rangle_{AB}\langle\Phi^2| + |\Phi^3\rangle_{AB}\langle\Phi^3|].
\end{equation}
The above equally weighted mixture has
\begin{equation}
{\cal C}[\chi_{AB}] = \frac{1}{2}(1 - \gamma) \not= 0
\end{equation}
unless $\gamma = 1$, but is useless for quantum teleportation since
\begin{equation}
{\cal F}[\chi_{AB}] = \frac{1}{2}.
\end{equation}
\item{(c)} $\eta^2 + \gamma^2 > 1$:
\begin{eqnarray}
\chi_{AB} & = & 
\frac{1}{Z}[e^{\beta{\cal B}}|\Phi^3\rangle_{AB}\langle\Phi^3| + e^{\beta J}|\Phi^2\rangle_{AB}\langle\Phi^2|] \nonumber \\
& \longrightarrow & |\Phi^3\rangle_{AB}\langle\Phi^3|,
\end{eqnarray}
where $Z$ is as in Eq.(13).  We have
\begin{eqnarray}
{\cal C}[\chi_{AB}] & = & \frac{\gamma}{\sqrt{\eta^2 + \gamma^2}}, \\
{\cal F}[\chi_{AB}] & = & \frac{1}{2}\left(1 + \frac{\gamma}{\sqrt{\eta^2 + \gamma^2}}\right) > \frac{1}{2}.
\end{eqnarray}
\end{description}
Therefore, $\eta_{\rm critical} = \sqrt{1 - \gamma^2}$ marks the point of quantum phase transition (phase transition taking place at zero temperature due to variation of interaction terms in the Hamiltonian of a system \cite{Arnesen}) from the maximally entangled state Eq.(13), which yields ${\cal C}[\chi_{AB}] = 1$ and ${\cal F}[\chi_{AB}] = 1$, to the generally nonmaximally entangled state Eq.(17), which still yields nonzero ${\cal C}[\chi_{AB}]$ and ${\cal F}[\chi_{AB}] > \frac{1}{2}$ as long as $\gamma$ is nonzero and $\eta$ finite: In the limit of large $\eta$,
\begin{equation}
{\cal C}[\chi_{AB}]\approx \gamma\eta^{-1} \longrightarrow 0,
\end{equation}\begin{equation}
{\cal F}[\chi_{AB}] \approx \frac{1}{2} + \frac{1}{2}\gamma\eta^{-1} \longrightarrow \frac{1}{2}
\end{equation}
when $\eta$ is infinitely large.  From Eqs.(15) and (18), we observe that for values of $\gamma$ other than $\gamma = \frac{1}{3}$, there is a sudden increase or decrease in ${\cal C}[\chi_{AB}]$ at $\eta_{\rm critical}$, depending on whether $\gamma > \frac{1}{3}$ or $\gamma < \frac{1}{3}$, before decreasing to zero asymptotically, as $\eta$ is increased beyond the critical value $\eta_{\rm critical}$.  In contrast, ${\cal F}[\chi_{AB}]$ always increases with increasing $\gamma$ at $\eta_{\rm critical}$ [see Eqs.(16) and (19)].

For nonzero temperatures, due to mixing of states, ${\cal C}[\chi_{AB}]$ decreases to zero and ${\cal F}[\chi_{AB}]$ decreases to $\frac{1}{2}$ as the temperature $T$ is increased beyond the critical values $T^{(1)}_{\rm critical}$ and $T^{(2)}_{\rm critical}$ respectively.  Except when $\gamma = 0$, both $T^{(1)}_{\rm critical}$ and $T^{(2)}_{\rm critical}$ depend on $\eta$.  If $\gamma = 0$, $T^{(1)}_{\rm critical} = 1.13459J$ for all $\eta$ \cite{Wang1}, but $T^{(2)}_{\rm critical}$ decreases from $1.13459J$ to zero as $\eta$ is increased from $0$ to $\eta_{\rm critical} = 1$, and remains zero when $\eta$ is increased beyond $1$ \cite{Yeo1}.  For nonzero $\gamma$, the behavior of $T^{(1)}_{\rm critical}$ and $T^{(2)}_{\rm critical}$ are qualitatively similar \cite{Yeo3}.  They both decrease with $\eta$ when $\eta$ is increased from $0$ to $\eta_{\rm critical} = \sqrt{1 - \gamma^2}$.  However, in this case, this is followed by monotonic increase as $\eta$ is increased beyond $\sqrt{1 - \gamma^2}$.  Indeed, in the limit of large $\eta$,
\begin{equation}
T^{(2)}_{\rm critical} \approx \frac{\eta J}{\ln\eta - \ln\gamma + \ln2}.
\end{equation}
Anisotropy therefore not only allows one to obtain entangled qubits at finitely higher $T$ and finitely higher $B_m$ than is possible in the isotropic case \cite{Kamta}, but also the associated entanglement is useful as resource for teleportation \cite{Yeo3}.

\section{The entanglement swapping scheme of Zukowski {\em et al.}}
Zukowski {\it et al.} \cite{Zukowski} presented an explicit entanglement swapping scheme to generate a three-qubit GHZ state \cite{Greenberger} from three Bell pairs:$A_1B_1$, $A_2B_2$, $A_3B_3$.  In their scheme, a local joint measurement in the three-qubit GHZ basis:
\begin{eqnarray}
|\Psi^0_{GHZ}\rangle & = & \frac{1}{\sqrt{2}}(|000\rangle + |111\rangle), \nonumber \\
|\Psi^1_{GHZ}\rangle & = & \frac{1}{\sqrt{2}}(|001\rangle + |110\rangle), \nonumber \\
|\Psi^2_{GHZ}\rangle & = & \frac{1}{\sqrt{2}}(|010\rangle + |101\rangle), \nonumber \\
|\Psi^3_{GHZ}\rangle & = & \frac{1}{\sqrt{2}}(|011\rangle + |100\rangle), \nonumber \\
|\Psi^4_{GHZ}\rangle & = & \frac{1}{\sqrt{2}}(|011\rangle - |100\rangle), \nonumber \\
|\Psi^5_{GHZ}\rangle & = & \frac{1}{\sqrt{2}}(|010\rangle - |101\rangle), \nonumber \\
|\Psi^6_{GHZ}\rangle & = & \frac{1}{\sqrt{2}}(|001\rangle - |110\rangle), \nonumber \\
|\Psi^7_{GHZ}\rangle & = & \frac{1}{\sqrt{2}}(|000\rangle - |111\rangle),
\end{eqnarray}
is performed on qubits $A_1$, $A_2$ and $A_3$ from each of the three Bell pairs.  They showed that this yields a GHZ state involving the other three qubits $B_1$, $B_2$ and $B_3$.  Here, instead of Bell pairs, we consider $\chi_{AB}$, Eq.(7).  The result is a three-qubit thermal state described by
\begin{equation}
\chi_{B_1B_2B_3} = \sum^7_{i = 1}p_i\chi^{(i)}_{B_1B_2B_3},
\end{equation}
where
\begin{equation}
\chi^{(i)}_{B_1B_2B_3} = \frac{1}{p_i}
{\rm tr}_{A_1A_2A_3}[(\Pi^i_{A_1A_2A_3} \otimes I_{B_1B_2B_3})(\chi_{A_1B_1} \otimes \chi_{A_2B_2} \otimes \chi_{A_3B_3})].
\end{equation}
The GHZ measurement is descibed by
\begin{equation}
\Pi^i_{A_1A_2A_3} \equiv |\Psi^i_{GHZ}\rangle_{A_1A_2A_3}\langle\Psi^i_{GHZ}|,\ i = 0, 1, 2, \cdots, 7,
\end{equation}
and
\begin{equation}
p_i \equiv {\rm tr}[(\Pi^i_{A_1A_2A_3} \otimes I_{B_1B_2B_3})(\chi_{A_1B_1} \otimes \chi_{A_2B_2} \otimes \chi_{A_3B_3})]
\end{equation}
is the probability that the measurement outcome $i$ occurs.  The entanglement swapping scheme of Zukowski {\it et al.} was subsequently generalized by Bose {\em et al.} \cite{Bose}, to the case of starting with $N$ different multipartite GHZ states involving any number of qubits, doing local joint multipartite GHZ basis measurements by selecting any number of qubits from the $N$ different multipartite GHZ states and also ending up with multipartite GHZ states involving any number of qubits.

\section{The teleportation scheme of Karlsson {\em et al.}}
In Ref.\cite{Karlsson} Karlsson {\it et al.} presented the teleportation of a two-level state $|\phi\rangle_{A_1}$ using a three-qubit GHZ state, for instance $|\Psi^0_{GHZ}\rangle_{A_2BC}$, to either $B$ or $C$ in such a way that, only one of them, say $C$, can fully reconstruct the quantum state conditioned on the measurement outcomes of the joint Bell measurement on $A_1$ and $A_2$:
\begin{equation}
\Pi^j_{A_1A_2} \equiv |\Psi^i_{Bell}\rangle_{A_1A_2}\langle\Psi^i_{Bell}|,\ j = 0, 1, 2, 3,
\end{equation}
and the single-qubit measurement on $B$:
\begin{equation}
\Pi^k_B \equiv |\psi^k\rangle_B\langle\psi^k|,\ k = 1, 2,
\end{equation}
with 
$|\psi^1\rangle_B \equiv \cos\mu|0\rangle_B + \sin\mu|1\rangle_B$ 
and 
$|\psi^2\rangle_B \equiv -\sin\mu|0\rangle_B + \cos\mu|1\rangle_B$ ($0 \leq \mu \leq \pi/4$).  Alternatively, the single-qubit measurement could be performed on $C$ such that $B$ reconstructs the quantum state.

In this paper, reliability for teleportation will be the criterion for judging the quality of the resulting three-qubit state, Eq.(24).  Therefore, in place of $|\Psi^0_{GHZ}\rangle_{A_2BC}$ we have $\chi^{(i)}_{A_2BC}$ [Eq.(25)].  The resulting state of qubit $C$ after the measurements is then given by
\begin{equation}
\rho^{(ijk)}_C = \frac{1}{q_{jk}}
{\rm tr}_{A_1A_2B}[(\Pi^j_{A_1A_2} \otimes \Pi^k_B \otimes I_C)(|\phi\rangle_{A_1}\langle\phi| \otimes \chi^{(i)}_{A_2BC})],
\end{equation}
where
\begin{equation}
q_{jk} \equiv {\rm tr}[(\Pi^j_{A_1A_2} \otimes \Pi^k_B \otimes I_C)(|\phi\rangle_{A_1}\langle\phi| \otimes \chi^{(i)}_{A_2BC})]
\end{equation}
is the probability that the measurement outcomes $j$ and $k$ occur.  The teleportation protocol is completed by applying to $\rho^{(ijk)}_C$ an appropriate Pauli rotation, which depending on $j$ and $k$, maximizes the teleportation fidelity:
\begin{equation}
\Phi_{\max} \equiv \frac{1}{4\pi}\int^{\pi}_0\sin\vartheta d\vartheta\int^{2\pi}_0d\varphi\ 
\sum^7_{i = 0}\sum^3_{j = 0}\sum^2_{k = 1}p_iq_{jk}\times {_C}\langle\phi|\rho^{(ijk)}_C|\phi\rangle_C,
\end{equation}
where the integration is over all pure two-level states, $|\phi\rangle = \cos\frac{\vartheta}{2}|0\rangle + e^{i\varphi}\sin\frac{\vartheta}{2}|1\rangle$ ($0 \leq \vartheta \leq \pi$, $0 \leq \varphi \leq 2\pi$).  $\Phi_{\max}$ quantitatively measures the reliability  of $\chi_{A_2BC}$ as resource for the  teleportation scheme of Karlsson {\em et al.}.

\section{Results}
For the two-qubit Heisenberg $XY$ chains at finite temperatures [Eq.(7)], we have
\begin{equation}
\Phi_{\max} = C_1 + C_2\cos\mu\sin\mu,
\end{equation}
where
\begin{equation}
C_1 = \frac{2(\cosh^2\beta{\cal B} + \cosh\beta{\cal B}\cosh\beta J + \cosh^2\beta J)}{3(\cosh\beta{\cal B} + \cosh\beta J)^2},
\end{equation}
\begin{equation}
C_2 = \frac{2(\sinh^3\beta J + \frac{\gamma J}{\cal B}\sinh^2\beta J\sinh\beta{\cal B} + \frac{\gamma^2J^2}{{\cal B}^2}\sinh\beta J\sinh^2\beta{\cal B} + \frac{\gamma^3J^3}{{\cal B}^3}\sinh^3\beta{\cal B})}{3(\cosh\beta{\cal B} + \cosh\beta J)^3}
\end{equation}
for both $\eta^2 + \gamma^2 \leq 1$ and $\eta^2 + \gamma^2 > 1$.  This is in contrast to ${\cal C}[\chi_{AB}]$ (see Ref.\cite{Kamta}) and Eq.(12).

In the zero temperature limit, Eq.(33) reduces to
\begin{equation}
\Phi_{\max} = \frac{2}{3}(1 + \cos\mu\sin\mu)
\end{equation}
when $\eta^2 + \gamma^2 < 1$, a signature of (pure) three-qubit GHZ teleportation-resource (see Ref.\cite{Karlsson});
\begin{equation}
\Phi_{\max} = \frac{1}{2} + \frac{1}{12}(1 + \gamma + \gamma^2 + \gamma^3)\cos\mu\sin\mu \leq \frac{2}{3}
\end{equation}
when $\eta^2 + \gamma^2 = 1$, in agreement with Eq.(16); and
\begin{equation}
\Phi_{\max} = \frac{2}{3}\left[1 + \left(\frac{\gamma}{\sqrt{\eta^2 + \gamma^2}}\right)^3\cos\mu\sin\mu\right]
\end{equation}
when $\eta^2 + \gamma^2 > 1$.  Therefore, $\Phi_{\max} > 2/3$ as long as $\gamma$ is nonzero and $\eta$ finite as in Eq.(21): In the limit of large $\eta$,
\begin{equation}
\Phi_{\max} \approx \frac{2}{3}(1 + \gamma^3\eta^{-3}\cos\mu\sin\mu) > \frac{2}{3},
\end{equation}
equals $2/3$ asymptotically when $\eta$ is infinitely large.

For nonzero temperatures, we observe that both $C_1$ and $C_2$ are positive definite.  Hence, for optimal $\Phi_{\max}$, we choose $\mu = \pi/4$ from hereon.  Due to mixing of states, $\Phi_{\max}$ decreases to $2/3$ at the critical temperature $T^{(3)}_{\rm critical}$ beyond which the teleportation fidelity is worse than what classical communication protocol can offer.  For each $\gamma$, $T^{(3)}_{\rm critical}$ is dependent on $\eta$.  To obtain $T^{(3)}_{\rm critical}$ we consider the following.  For the thermal state Eq.(24) to be useful for quantum teleportation at nonzero $T$, we demand that
\begin{eqnarray}
& & \sinh^3\beta J + \frac{\gamma}{\sqrt{\eta^2 + \gamma^2}}\sinh^2\beta J\sinh\beta\sqrt{\eta^2 + \gamma^2}J \nonumber \\
& & + \left(\frac{\gamma}{\sqrt{\eta^2 + \gamma^2}}\right)^2\sinh\beta J\sinh^2\beta\sqrt{\eta^2 + \gamma^2}J 
+ \left(\frac{\gamma}{\sqrt{\eta^2 + \gamma^2}}\right)^3\sinh^3\beta\sqrt{\eta^2 + \gamma^2}J \nonumber \\
& > & 2(\cosh^2\beta J\cosh\beta\sqrt{\eta^2 + \gamma^2}J + \cosh\beta J\cosh^2\beta\sqrt{\eta^2 + \gamma^2}J).
\end{eqnarray}

\begin{table}
\begin{tabular}{|l|l|l|l|l|l|l|l|l|l|l|}
\hline
$\eta$                     & 0 & 0.1 & 0.2 & 0.3 & 0.4 & 0.5 & 0.6 & 0.7 & 0.8 & 0.9 \\ \hline
$T^{(2)}_{\rm critical}/J$ & 1.13459 & 1.13105 & 1.12029 & 1.10193 & 1.07525 & 1.03904 & 0.99126 & 0.92828 & 0.84267 & 0.71411 \\ 
\hline
$T^{(3)}_{\rm critical}/J$ & 0.55508 & 0.55093 & 0.53825 & 0.51631 & 0.48371 & 0.43810 & 0.37605 & 0.29473 & 0.19890 & 0.09950 \\ 
\hline
\end{tabular}\\[1ex]
Table 1: Critical temperatures $T^{(2)}_{\rm critical}$ and $T^{(3)}_{\rm critical}$ for various values of $\eta$, when $\gamma = 0$.
\end{table}

When $\gamma = 0$, Eq.(40) reduces to
\begin{equation}
\sinh^3\beta J > 2(\cosh^2\beta J\cosh\beta\eta J + \cosh\beta J\cosh^2\beta\eta J).
\end{equation}
Hence, $T^{(3)}_{\rm critical}$ can be obtained from
\begin{equation}
\sinh^3\frac{J}{T^{(3)}_{\rm critical}} = 
2\left(\cosh^2\frac{J}{T^{(3)}_{\rm critical}}\cosh\frac{\eta J}{T^{(3)}_{\rm critical}} 
+ \cosh\frac{J}{T^{(3)}_{\rm critical}}\cosh^2\frac{\eta J}{T^{(3)}_{\rm critical}}\right).
\end{equation}
It decreases from $0.555081J$ to $0$ as $\eta$ is increased from $0$ to $\eta_{\rm critical} = 1$, and remains zero when $\eta$ is increased beyond $1$ (see FIG. 1).  This is because Eq.(41) is unattainable, consistent with the fact that the original two-qubit thermal states [Eq.(7)] have fully entangled fraction less or equal to $1/2$ and thus cannot yield nonclassical teleportation fidelity, when $\eta \geq 1$.  From Eq.(36) and a comparison of $T^{(3)}_{\rm critical}$'s with the corresponding $T^{(2)}_{\rm critical}$'s from Ref.\cite{Yeo1} (see Table 1), we conclude that the entanglement swapping process ``amplifies'' the thermal noise present in the original two-qubit states [Eq.(7)] resulting in $T^{(3)}_{\rm critical} < T^{(2)}_{\rm critical}$ for each given $\eta$.

For nonzero $\gamma$, Eq.(40) can be satisfied as long as $\eta^2 + \gamma^2 < 1$ or $\eta^2 + \gamma^2 > 1$.  At $\eta^2 + \gamma^2 = 1$, it reduces to
\begin{equation}
(1 + \gamma + \gamma^2 + \gamma^3)\sinh^3\beta J > 4\cosh^3\beta J,
\end{equation}
which is again unattainable.  Therefore, $T^{(3)}_{\rm critical}$ similarly decreases to zero when $\eta$ is increased from $0$ to $\eta_{\rm critical} = \sqrt{1 - \gamma^2}$, as shown in FIG. 1.  This is then followed by a monotonic increase in $T^{(3)}_{\rm critical}$ as $\eta$ is increased beyond $\sqrt{1 - \gamma^2}$.  Indeed, in the limit of large $\eta$:
\begin{equation}
T^{(3)}_{\rm critical} \approx \frac{\eta J}{3\ln\eta - 3\ln\gamma + \ln2},
\end{equation}
as in Eq.(22).  The behavior of $T^{(3)}_{\rm critical}$ is therefore qualitatively similar to $T^{(1)}_{\rm critical}$ and $T^{(2)}_{\rm critical}$ (compare FIG. 4 in Ref.\cite{Kamta} and FIG. 1 in Ref.\cite{Yeo3} with our FIG. 1).  Hence, anisotropy allows one to obtain three-qubit GHZ-like thermal states Eq.(24), which yield nonclassical teleportation fidelities at finitely higher temperatures by applying an external magnetic field of finitely greater strengths.  This is somewhat surprising in view of the following two facts.  First, even in the zero temperature limit, $|\Phi^3\rangle_{AB}$ is far from ``ideal'': Eqs.(17) - (21).  Second, from the above conclusion, one would naturally expect these imperfections to be amplified.

\begin{figure}
\caption{Temperature $T^{(3)}_{\rm critical}/J$ at which the teleportation fidelity is less than $2/3$, plotted as a function of $\eta$ (magnetic field strength $B_m$) for various values of the anisotropy parameter: $\gamma = 0$ (solid line), $\gamma = 0.3$ (dash-dotted line), $\gamma = 0.6$ (thick dashed line), and $\gamma=1$ (thin dashed line).  In each case, the teleportation fidelity is less than or equal to $2/3$ in the region bounded by (and generally above) the relevant curve. Note that for any finite temperature, there is an $\eta$ for which the fidelity is strictly greater than $2/3$.}
\end{figure}

\section{Conclusions}
In summary, we have shown that despite the fact our entanglement swapping procedure amplifies thermal noise, anisotropy in the original Heisenberg $XY$ chains enables the resulting three-qubit GHZ-like mixed state to yield nonclassical teleportation fidelities at finitely higher and higher temperatures by suitably increasing the strengths of an external magnetic field.  In the light of the result of Gottesman and Chuang \cite{Gottesman}, our result has therefore a positive implication on the quantum-dot realization of a quantum computer.   However, we should point out that although nonclassical, the teleportation fidelities are very close to $2/3$, as can easily be deduced from Eq.(39).  Hence, one immediate research problem would be to look at models like those considered by Anteneodo {\it et al.} \cite{Anteneodo} and Zhou {\it et al.} \cite{Zhou} to see if this situation could be improved.  Next, multi-qubit GHZ states would be necessary in any realistic setting.  To produce these states would require more ``complicated'' GHZ basis measurements.  Therefore, another interesting research problem would be to study how the noise amplification scales with the complexity of these measurements.


\begin{thebibliography}{99}
\bibitem{Comment} Thermal entanglement in three-qubit Heisenberg $XX$ and $XXZ$ models were being studied in Ref.\cite{Wang3}.  However, the entangled energy-eigenstates for these models belong to the class of $W$-states \cite{Dur}, which is fundamentally different from the class of GHZ states considered by Gottesman and Chuang in Ref.\cite{Gottesman}.
\end{thebibliography}
\end{document}